\documentclass[proceedings, preprint]{rmaa}

\usepackage{paralist}
\usepackage{psfrag,color}

\newcommand{\sr}{R_{\odot}}
\newcommand{\rsh}{{\partial \Omega/\partial r}}

\SetYear{2008}
\SetConfTitle{MFU II conference} 

\title{Flux transport solar dynamo models, current problems and
  possible solutions} 

\author{
  G. Guerrero,\altaffilmark{1}
  and E. M. de Gouveia Dal Pino \altaffilmark{1}
}

\altaffiltext{1}{Astronomy Department, Instituto de Astronomia,
  Geof\'{i}sica e Ci\^{e}ncias Atmosf\^{e}ricas,
  Universidade de S\~{a}o Paulo,
  Rua do Mat\~{a}o 1226, S\~{a}o Paulo, Brazil.\\
  (guerrero,dalpino@astro.iag.usp.br).}

\shortauthor{Guerrero \& de Gouveia Dal Pino}
\shorttitle{Solar dynamo, current problems and possible solutions}

\listofauthors{G. Guerrero \& E. M. de Gouveia Dal Pino}
\indexauthor{G. Guerrero}
\indexauthor{E. M. de Gouveia Dal Pino}

\abstract{The sunspot solar cycle has been usually explained as the
  result of a dynamo process operating in the sun. This is
  a classical problem in Astrophysics that until the present is not fully
  solved. Here we discuss  current problems and limitations with the
  solar dynamo modeling and their possible solutions using the
  kinematic dynamo model with the Babcock-Leighton 
  approximation as a tool. In particular,
  we discuss the importance of the turbulent
  magnetic pumping versus the meridional flow circulation in the
  dynamo operation.} 

\addkeyword{Sun: magnetic fields}

\begin{document}
\maketitle

\section{Introduction}
\label{sec:intro}

The sunspot cycle is one of the most interesting magnetic phenomenon
in the Universe. It was discovered more than 150 years ago by
\citet{schwabe}, but until now it remains  an open problem in
astrophysics. There are several large scale observed phenomena that
evidence that the solar cycle corresponds to a dynamo process
operating inside the sun. These can be summarized as follows:

\begin{inparaenum}[(1)]
\item The sunspots usually appear in pairs at both sides of the solar
  equator; the leading spot of a pair  (i.e. the one that points to the
  E-W direction) has opposite polarity to the other one.
  Besides, leading spots in the northern hemisphere
  have the opposite polarity to that of the leading spots in the southern
  hemisphere. Sunspots invert their polarity every 11 years; the total
  period of the cycle is then 22 years. This is known as the Hale's
  law; 
 \item A straight line connecting the leading and the companion spots
 of a pair has always an inclination of $10^{\circ}$ to $30^{\circ}$
   with respect to the equatorial line. This is known as the Joy's law;
 \item When the toroidal field reaches its maximum, i.e., when
   the number of sunspots is maximum, the global poloidal field
   inverts its polarity, so that there is a phase lag of
   $\pi/2$ between the toroidal and poloidal components of the
   magnetic field;
 \item The sunspots appear only in a belt of latitudes between $\pm
   30^{\circ}$ (at both sides of the solar equator), these are known as
   the latitudes of activity;
 \item The strength of the magnetic fields in the sunspots is around
 $10^3$ G. The magnitude of the diffuse poloidal  field is of tens of G.
\end{inparaenum}

\citet{parker55} was the first to try to explain the solar cycle as
a hydromagnetic phenomenon, since then although there has been
important improvements in the observations, theory and simulations,
a definitive model for the solar dynamo is still missing.
Helioseismology has mapped the solar internal rotation showing a
detailed profile of the latitudinal and radial shear layers, which
seems to confirm the usually accepted idea that the first part of
the dynamo process is the transformation of an initial poloidal
field into a toroidal one. This stage is known as the $\Omega$
effect. The second stage of the process, i.e., the transformation of
the toroidal field into a new poloidal field of opposite polarity is
a less understood process, and has been the subject of intense
debate and research. Two main hypotheses have been formulated in
order to explain the nature of this effect, usually denominated the
$\alpha$ effect: the first one is based on the Parker's idea of a
turbulent mechanism where the poloidal field results from cyclonic
convective motions operating at small scales in the toroidal field.
These small loops should reconnect to form a large scale dipolar
field. However, these models face an important problem: in
the non-linear regime, i.e. when the back reaction of the toroidal
field on the motions becomes important, the $\alpha$ effect can be
catastrophically quenched \citep{vainsh92} leading to an ineffective
dynamo \citep{cattaneo96}\footnote{Nevertheless, it is noteworthy that   
when the magnetic helicity is included in dynamical
computations of $\alpha$, it does not become catastrophycally quenched
as long as the flux of the magnetic helicity 
remains non null \citep[see][for a complete review of this
subject]{bs05}. The shear in the fluid could be the way
through which the helicity flows to outside of the domain 
\citep{vishniac}.}.

The second one is based on the formulations of
\citet{babcock} and \citet{leighton} (BL). They proposed that
the inclination observed in the bipolar magnetic regions (BMR's)
contains a net dipole moment. The supergranular diffusion causes the
drift of half of each of these active regions to the equator and the
drift of the other half in direction to the poles.
so that this large scale
poloidal structure annihilates the previous dipolar field. 
The new dipolar field is transported by the meridional circulations to 
the higher latitudes in order to form the observed polar field. 
This second mechanism has the
advantage of being directly observed at the surface
\citep{wangetal89,wangetal91}, but it does not 
discard the existence of other $\alpha$ sources underneath.

Following the BL idea, the physical model for the solar dynamo
begins with a dipolar field. The differential rotation stretches the
poloidal lines and form a belt of toroidal field at some place within
the solar interior, in the convective layer. This toroidal field is
somehow pushed through the 
turbulent convective eddies and forms strong and well organized
magnetic flux tubes. When the magnetic field is intense enough and
the density inside a tube is lower than the density of the
surrounding plasma, it becomes unstable and begins to emerge towards
the surface where it will form the BMRs. By diffusive decay, a BMR
will form a net dipolar component with the opposite orientation of
the original one, this new dipolar field is amplified during the cycle
evolution until it reverses the previous dipolar field. In order to
complete the 
cycle, it is necessary to transport this new poloidal flux first to
the poles and then to the
internal layers where the toroidal field will be created again, and
so on. Most of the models in the BL mechanism use the meridional
circulation  flow as the main agent of transport.  Recent works have
invoked the turbulent pumping as an additional mechanism to advect
the magnetic flux (see below).

In the absence of direct observations to confirm  the model above,
several numerical studies have been performed in order to simulate
the solar dynamo. These can be divided in two main classes: global
dynamical models \citep{brun04} and mean field kinematic models
\citep{dik99,ccn04,kuketal01,bon02,gue04,kapia06}. The first class
integrates the full set of MHD equations in the solar convection
zone and employ  the inelastic approximation in order to overcome the 
numerical constrain imposed by fully compressible convection on the
time-step. These models are able to reproduce the observed
differential rotation pattern, but they do not generate a cyclic, and
well organized pattern of toroidal magnetic field. 
The second class of simulations solves the
induction equation only and uses observed and/or estimated profiles
for the velocity field and the diffusion terms. These models are 
relatively successful in reproducing the large scale features of the
solar cycle, but the lack of the dynamical part of the problem has
led to uncertainties in  the dynamo mechanism.

We have carried out several numerical tests with a mean field dynamo
model in the Babcock-Leighton approximation in order to search for
answers to four main issues:

\begin{inparaenum}[(1)]
\item where is the solar dynamo located?;
\item what is the dominant flux transport mechanism?;
\item how to explain the observed latitudes of the solar activity?;  and
\item why is the solar parity  anti-symmetric?
\end{inparaenum}

In the following paragraphs, we briefly summarize our model
assumptions and, step by step, draw our approach to the questions
above in the light of the numerical simulations.

\section{Model}
\label{sec:model}

The equation that describes the temporal and spatial evolution of the
magnetic field is the induction equation:
\begin{equation}\label{eq1}
\frac{\partial {\bf B}}{\partial t} = \nabla \times [{\bf U} \times
    {\bf B} + {\bf \cal{E}} - \eta \nabla \times {\bf B} ],
\end{equation}
where ${\bf U}$$=$${\bf u_p} + {\bf \Omega} r \sin \theta$ is the
observed velocity field, $\Omega$ is the angular velocity, ${\bf
B}$$=$$\nabla \times (A {\bf \hat{e}_{\phi}}) + B_{\phi} {\bf
  \hat{e}_{\phi}}$, where $\nabla \times (A {\bf \hat{e}_{\phi}})$ and
$B_{\phi}$ are the poloidal and 
toroidal components of the magnetic field, respectively,
$\eta$ is the microscopic magnetic diffusivity and
\begin{eqnarray}\label{eq2}
{\bf \cal{E}} &=& {\bf \alpha} {\bf B} + {\bf \gamma} \times {\bf B} -
\beta (\nabla \times {\bf B})\\\nonumber 
&-& \delta \times (\nabla \times {\bf B}) - \kappa (\nabla  {\bf B}) \quad,  
\end{eqnarray}
corresponds to the first and second order terms of the expansion of
the electromotive force, $\overline{{\bf u} \times {\bf b}}$, and
represents the action of the small-scale fluctuations over the large
scales. The coefficients of (\ref{eq2})  are the so-called dynamo
coefficients. The first term on the right hand side of eq. (2)
corresponds to the turbulent $\alpha$ effect coefficient, not
considered in our Babcock-Leighton formulation. The second one is the
turbulent magnetic pumping. The third corresponds to the turbulent 
diffusivity, which in our model is combined with the microscopic value 
($\eta_T$$=$$\eta + \beta$). For the sake of simplicity, the other two
terms are neglected. We solve  equation (\ref{eq1}) for $A$ and
$B_{\phi}$ with $r$ and $\theta$ coordinates in the spatial ranges
$0.6\sr$$-$$\sr$ and $0$$-$$\pi$, respectively, in a $200 \times 200$
grid esolution (see \citet{gue04}, for details regarding the numerical  
model). 

\begin{figure}[!t]
  \includegraphics[width=\columnwidth]{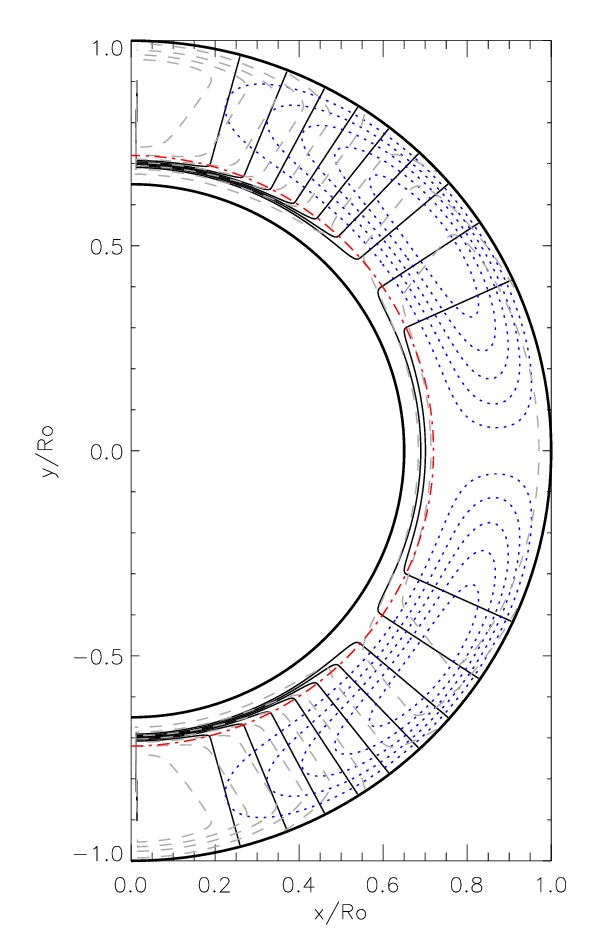}
  \caption{Isorotation lines of the solar angular velocity as inferred
     from helioseismology observations (continuous lines) and the
     adopted meridional circulation stream lines. The red dot-dashed
     line shows the boundary between the overshoot layer and the
     convection zone.}
  \label{fig:1}
\end{figure}

The profiles that we employ describe the results of recent
helioseismology inversions or numerical simulations. For the
differential rotation, we consider a profile mapped from
helioseismology (see the 
continuous lines of Figure 1). For the meridional flow, we consider
one cell per meridional quadrant, as usually assumed (doted lines in
Figure 1). The alpha term ($\alpha B_{\phi}$) is concentrated
between $0.95 \sr$ and $\sr$ and at the latitudes where the sunspots
appear (see the continuous lines in Figure 2). Since it must result
the emergence of magnetic flux tubes, we assume this term to be
proportional to the toroidal field $B_{\phi}(r_c,\theta)$ at the
overshoot interface between the radiative and the convective
regions, $r_c$$=$$0.715\sr$. For the magnetic diffusion, we consider
only one gradient of diffusivity located at $r_c$ which separates
the radiative stable region (with $\eta_{rz}$$=$$10^9$ cm s$^{-2}$)
from the convective turbulent layer (with $\eta_{cz}$$=$$10^{11}$ cm
s$^{-2}$) (see the dotted line in the upper panel of Figure 2).

\begin{figure}[ht]
\centering
  \includegraphics[width=\columnwidth]{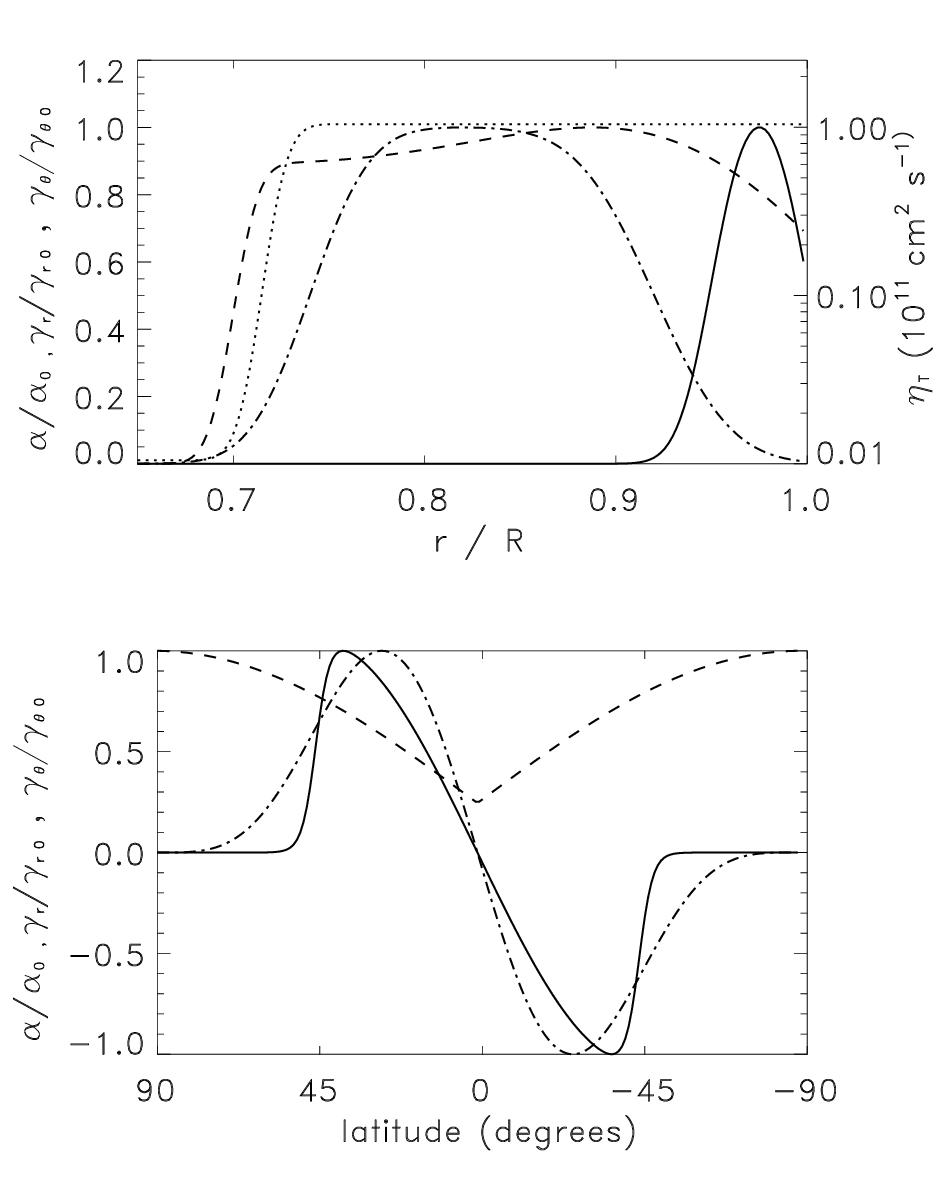}
  \caption{Radial and latitudinal profiles for $\alpha$ (continuous
  line), $\eta_T$ (dotted line) and for the pumping terms $\gamma_r$ and
  $\gamma_{\theta}$ (dashed and dot-dashed lines,
  respectively). All the profiles are normalized to their maximum
  value. {\it Extracted from \citet{gue08}}.}
  \label{fig:2}
\end{figure}

\section{The location of the solar dynamo}

As remarked before, the differential rotation pattern which is
responsible for the $\Omega$ effect, is revealed by high
resolution helioseismology observations. It describes a solid-body
rotation for the 
radiative core, and a differentially rotating convective layer
with a retrograde velocity with respect to the
radiative interior at higher latitudes and a pro-grade velocity at
lower latitudes. The interface that bounds the solid-body rotation
zone is named tachocline $-$ its exact location and width have not
been  established yet. Another radial shear layer has been recently
identified just below the solar photosphere in the upper $35$ Mm of
the sun \citep{cor02} (see the gray dashed line in Fig. 1). With
this newly discovered shear layer, it is even more difficult to
define where the dynamo operates. There has been so far, an apparent
common agreement that the dynamo is operating at the tachocline.
However this possibility has several problems \citep{bran05}. One of
the main difficulties is that toroidal flux ropes formed in the
tachocline should have intensities  $\sim$ $10^4$-$10^5$ G in order
to become buoyantly unstable and to emerge at the surface to form a
BMR the appropriate tilt given by the
Joy's law
\citep{dsilva93,fanetal93,caligetal95,caligetal98,fanfisher96,fan04}.
One important limitation of this scenario is that $10^5$ G results
an energy density that is an order of magnitude larger than the
equipartition value, so that a stable layer is required to store and
amplify  this magnetic field. This raises another question with
regard to the way in which the magnetic flux is dragged down to
deeper layers. In recent work \citep{gue07a}, we have explored the
contributions of the shear terms in the dynamo equation, $({\bf B_p}
\cdot \nabla) \Omega$$=$$B_r\rsh + B_{\theta}/r \partial \Omega
\partial \theta$ with the aim of determining where the most strong
toroidal magnetic fields are produced. We found that the radial
shear component is about two orders of magnitude smaller than the
latitudinal component. Therefore, when a new toroidal field begins
to develop its growth is dominated by the latitudinal shear. Its
amplification begins in the bulk of the convection zone and it is
transported to the stable layer where it must reach the desired
magnitude (see Fig. 3). How this flux is transported is the subject
of the next section.

\begin{figure}[!t]
\centering
  \includegraphics[width=\columnwidth]{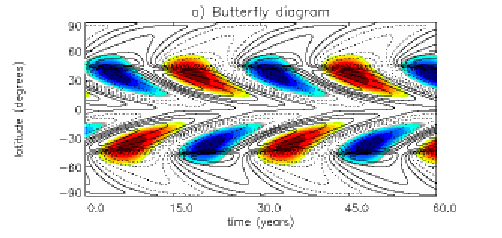}\\
  \includegraphics[width=\columnwidth]{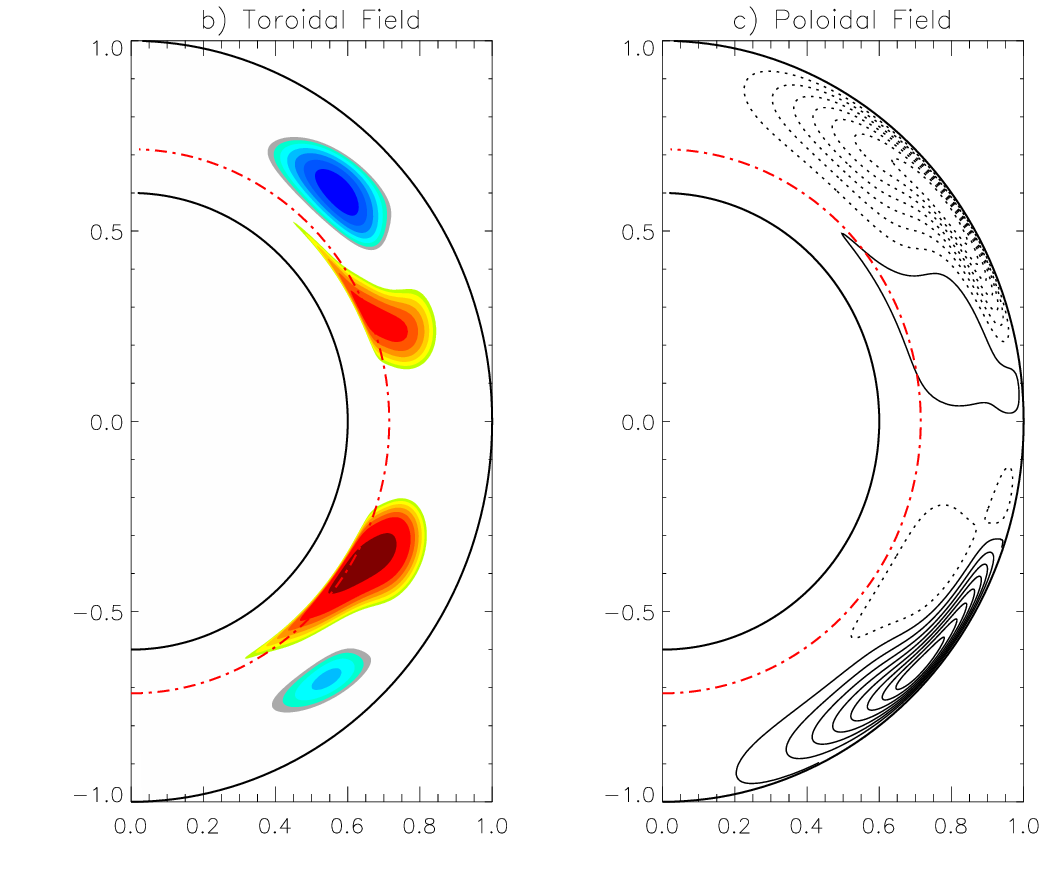}
\caption{(a) Butterfly diagram and latitudinal
  snapshots for the toroidal (b) and the poloidal (c) fields. The dark
  (blue) and light (red) gray (color) scales represent positive and
  negative toroidal fields, respectively; the continuous and dashed
  lines represent the positive and negative poloidal fields. Only
  toroidal fields greater than $2\times10^4$ G (the most external
  contours) are shown in panels (a) and (b). This model started with
  an anti-symmetric initial condition (see \S6 for details). {\it
  Extracted from \citet{gue08}}.}\label{fig3}
\end{figure}

If a near-surface shear layer is turned on in the model (Guerrero \&
de Gouveia Dal Pino 2008 and references therein), two main branches
appear in the butterfly diagram (Fig. 4). One is migrating poleward
(at high latitudes) and another is migrating equatorward (below
$45^{\circ}$). This result is expected if the Parker-Yoshimura sign
rule \citep{parker55,yoshi75} applies. This contribution to the
toroidal field was previously explored in a Babcock-Leighton dynamo
by \citet{diketal02}. They have discarded the near-surface radial
shear layer because it generates butterfly diagrams in which a
positive toroidal field gives rise to a negative radial field, which
is exactly the opposite to the observed. Our results, on the other
hand, present the 
correct phase lag between the fields. This difference probably
arises from the fact that we are using a lower meridional
circulation amplitude.  Anyway, as can be seen in Fig. 4, the polar
branches are strong enough to generate also undesirable sunspots close
to the poles. The period increases to $15.6$ y due to the fact that
the dominant dynamo action at the surface goes in the opposite
direction to the meridional flow (i.e, the dynamo wave direction is
dominating over the meridional flow).

\begin{figure}[!t]
\centering
  \includegraphics[width=\columnwidth]{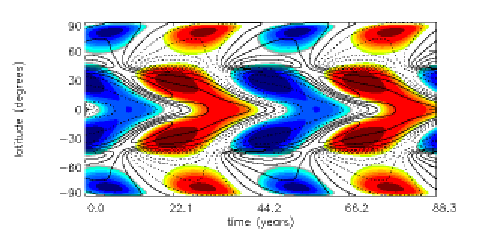}
\caption{Butterfly diagram for a model with the same
  parameters as in Fig. 3, but with near-surface shear layer. For this
  model $T$$=$$15.6$ yr $B_{\phi_{max}}(r$$=$$0.715)$$=$$1.1 \times
  10^5$,  $B_{\phi_{max}}(r$$=$$0.98)$$=$$1.9 \times 10^4$  G and
  $B_{r_{max}}$$=$$131.7$ G. This model started with
  symmetric initial conditions. {\it Extracted from
  \citet{gue08}}.}\label{fig6}
\end{figure}

\section{Flux transport mechanisms}

For an $\alpha \Omega$ dynamo, the Parker-Yoshimura sign rule
\citep{parker55,yoshi75} establishes that the direction of the
dynamo wave is equatorward or poleward if the product $\alpha$
$\cdot$ $\rsh$ is $<$ or $>$$0$, respectively. Hence, models
with a solar like rotation law operating at the tachocline, with a
positive $\alpha$ effect in the northern hemisphere, as believed,
and without meridional circulation, should result a solution with
magnetic branches migrating in the opposite way to that observed
\citep{kuketal01}. Models with one cell of meridional circulation,
poleward at the surface and equatorward at the base of the
convection zone, produce results with the appropriate direction of
propagation \citep{dik99,kuketal01,bon02}. It was demonstrated also
that the meridional circulation sets the period of the cycle
\citep{dik99} and that it is the most logical way to transport the
novel poloidal fields to the inner layers in the dynamo process.
These models in which the time of the cycle fits better with the
advective time than with the diffusive one are usually called
advection dominated or flux-transport dynamos. There is, however, an
important problem with these models: they require a large scale
meridional flow and this is observed only at the surface. It is
possible that the real meridional flow is too weak at the inner
regions to penetrate the overshoot layer and the tachocline
\citep{gil04,rudetal05}, or perhaps, it has a multicell pattern
\citep{mitra07,bon06,jouve07}. If this is the case, it is very hard
to explain the equatorward migration of the toroidal branches and it
is necessary to find another flux-transport mechanism. The turbulent
pumping seems to be a good candidate.

The turbulent pumping effect corresponds to the transport, in all
directions,  of magnetic flux due to the presence of density
(buoyancy) and turbulence (diamagnetism) gradients in convectively
unstable layers. In the FOSA \citep[First Order Smoothing
  Approximation, see e.g.][and references therein]{bransub}, the
radial component of the diamagnetic pumping 
can be calculated assuming a linear dependence with the variations in
the magnetic diffusivity $U_{dia}$$=$$-\nabla \eta_T /2$; the buoyancy
component depends on the density gradients $U_{buo}$$=$$-\eta \nabla
\rho / \rho$. In the
boundary between the solar convection zone and the overshoot layer,
it is probable that the diamagnetic velocity is of the order of $50$
ms$^{-1}$ \citep{kitrud}. This value strengthens the importance of
the pumping relative to the assumed radial meridional flow velocity
which is $<10$ ms$^{-1}$. For our model, we obtain
$U_{dia}$$=$$\simeq$$47$ cm s$^{-1}$ when a variation of two orders
of magnitude is considered in the diffusivity in a thin region of
$0.015\sr$ \citep{gue08} (the buoyancy component of the pumping is not
considered because the density is not a parameter in kinematic
models). 

\begin{figure}[!t]
\centering
  \includegraphics[width=5cm]{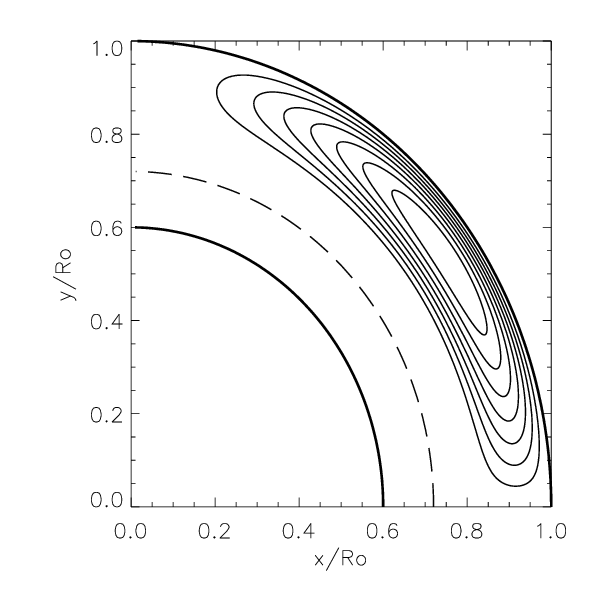}\\
  \includegraphics[width=\columnwidth]{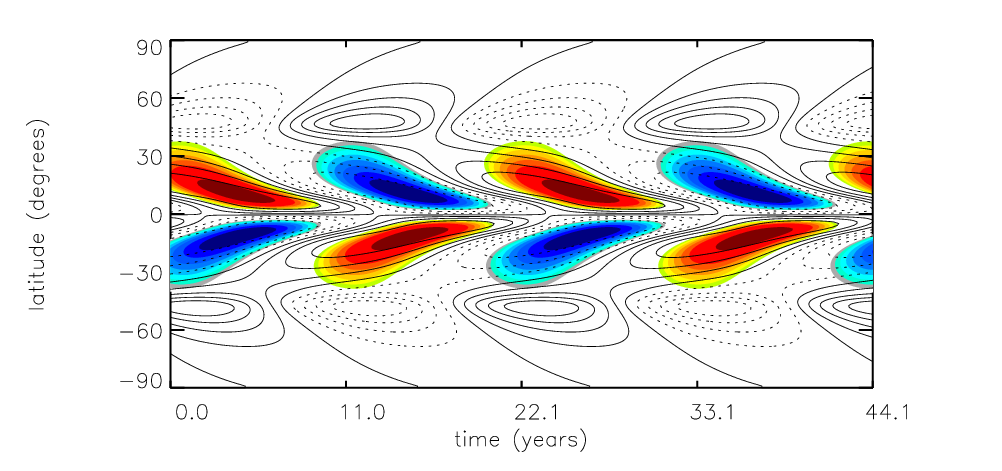}
  \includegraphics[width=\columnwidth]{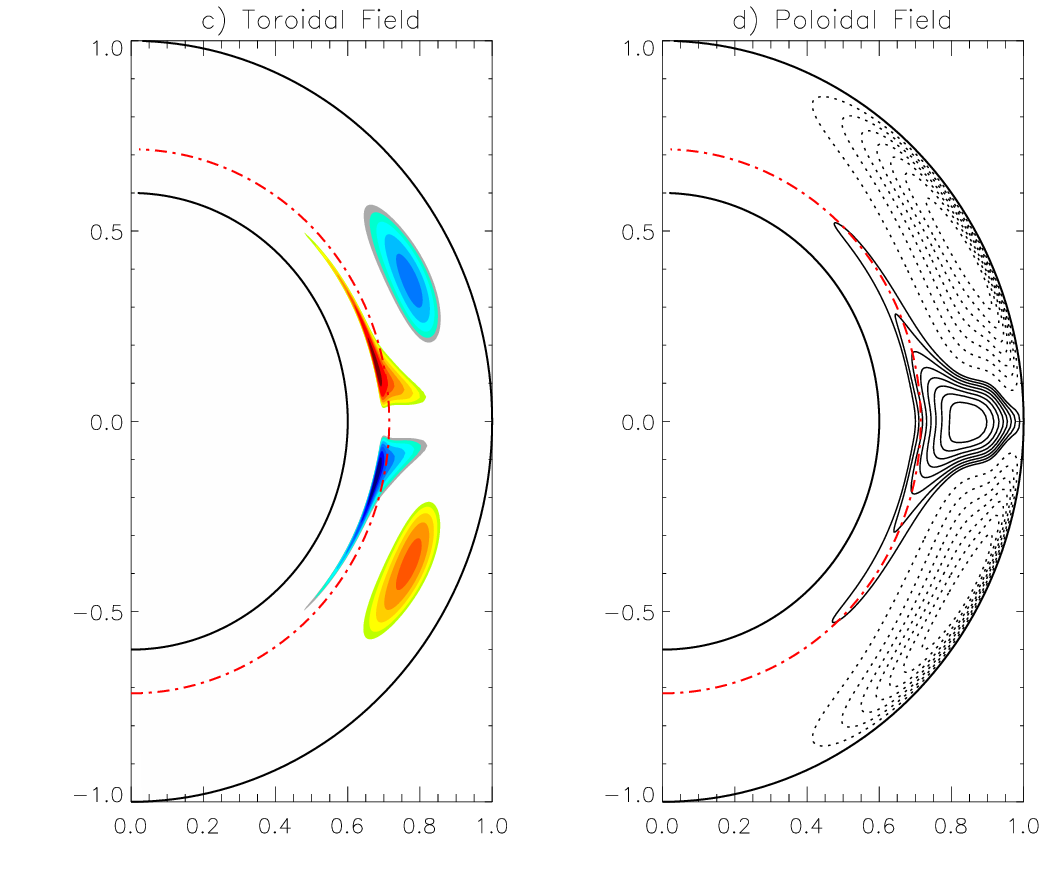}
\caption{Meridional flow streamlines and the butterfly diagram for a
model with the full pumping term, but with a shallow meridional flow
penetration with a depth of only $0.8\sr$, $U_0$$=$$1300$ cm
s$^{-1}$, $\gamma_{\theta 0}$$=$$90$ cm s$^{-1}$ and $\gamma_{r
0}$$=$$30$ cm s$^{-1}$. For this model we obtain $T$$=$$10.8$ yr,
  $B_{\phi_{max}}$$=$$4.5 \times 10^4$ G and $B_{r_{max}}$$=$$154.9$
  G. This model started with anti-symmetric initial
  conditions. {\it Extracted from \citet{gue08}}.}\label{fig5}
\end{figure}

The effects of turbulent pumping have been rarely considered in mean
field dynamo models. A first approach showing its importance in the
solar cycle was made by \citet{branetal92}; since then few works
have incorporated the diamagnetic pumping component in the dynamo
equation as an extra diffusive term that can provide a downward
velocity, as discussed above \citep{kuketal01,bon02,bon06}. More
recently, \citet{kapia06} have implemented simulations of the mean
field dynamo in the distributed regime, including all the dynamo
coefficients previously evaluated in magneto-convection simulations
\citep{ossetal02,kapia06a}. They produced butterfly diagrams that
resemble the observations. However, to our knowledge no special
efforts have been made to study the pumping  effects in the
meridional plane (i.e., inside the convection zone) or in a BL
description.

We have included the turbulent pumping terms (see the
dashed and dot-dashed lines in Fig. 2 which correspond to the radial,
$\gamma_r$, and latitudinal, $\gamma_{\theta}$, pumping terms)
calculated from local magneto-convection simulations
\citep{ossetal02,kapia06a} into the induction equation (eq. 1).

For a dynamo model operating at the tachocline, we find
that the pumping terms lead to a distinct
latitudinal distribution  of the toroidal fields when compared with
the results of Fig. 3.  The turbulent and density
gradient levels present in the convectively unstable layer cause
the pumping of the magnetic field both down and equartorward, allowing
its amplification within the stable layer and its later emergence at
latitudes very near the equator.

If we also include in this model recent helioseismic results
\citep{mitra07} that suggest that the return point of the meridional
circulation can be at $\sim 0.95\sr$, at lower regions, beneath
$\sim 0.8\sr$, a second weaker convection cell or even a null large
scale meridional flow can exist. In Figure 5, we obtain a butterfly
diagram that agrees with the main features of the solar cycle,
besides, we find that in this case, it is the pumping terms that
regulate the period of the cycle, leading to a $different$ $class$
of dynamo that  is advection-dominated  by turbulent pumping rather
than by a deep meridional flow.

For the dynamo model operating at the near-surface shear layer, we
find that the toroidal fields created at high latitudes are
efficiently pushed down before reaching a significant amplitude, so
that only the equatorial branches below $45^{\circ}$ survive (see
Fig. 6). This result is explained by the fact that the radial
pumping component has its maximum amplitude close to the poles (see
the dashed line in Fig. 2). These results also agree with the
observations.

\begin{figure}[!t]
\centering
  \includegraphics[width=\columnwidth]{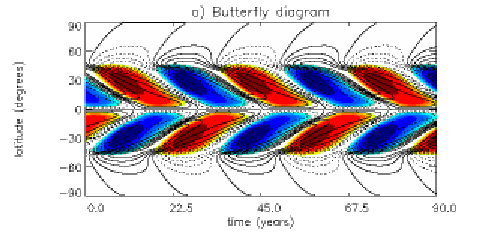}\\
  \includegraphics[width=\columnwidth]{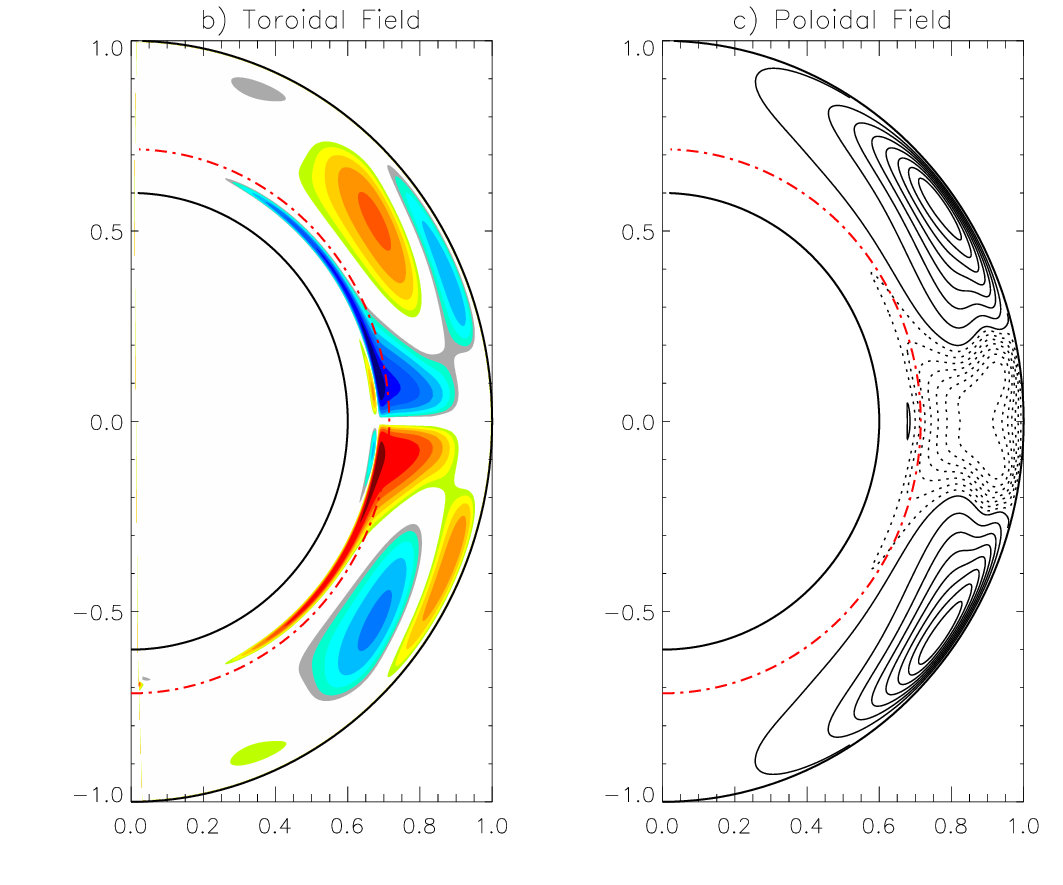}
\caption{The same as in Fig. 5, but for a model with
  near-surface shear
  action. For this model $T$$=$$16.3$ yr
  $B_{\phi_{max}}(r$$=$$0.715)$$=$$9.7 \times 10^4$,
  $B_{\phi_{max}}(r$$=$$0.98)$$=$$1.9 \times 10^4$  G and
  $B_{r_{max}}$$=$$164.4$ G. This model started with
  symmetric initial conditions. {\it Extracted from
  \citet{gue08}}.}\label{fig6}
\end{figure}

\section{How to explain the observed latitudes of the solar activity?} 

The radial shear at the tachocline has its maximum amplitude in
regions close to the poles, for this reason, it is a common problem
in  mean field dynamo models to present large undesirable toroidal
magnetic fields in the polar regions. \citet{nandy02} proposed a
deep meridional flow as a way to avoid the formation of strong
toroidal fields at high latitudes. Under this assumption, the strong
toroidal magnetic fields formed at high latitudes are pushed down,
inside the radiative zone, by the meridional flow and are stored
there until they reach latitudes below $30^{\circ}$. However this
assumption may lead to undesirable mixing of the chemical elements
between the radiative and convective zones and to problems regarding 
the angular momentum transfer. Besides, some results of numerical
simulations show that the meridional flow is unable to penetrate the
tachocline \citep{gil04,rudetal05}.

We have explored this subject in two ways. First, we built a hybrid
model in which we combined the profiles used by \citet{nandy02} with
those used by \citet{dik99} and allowed a deep meridional flow
\citep{gue04}. We found that the high-latitude toroidal field is
sensitive to the model. Then, we explored this problem by changing
the shape and the thickness  of the solar tachocline \citep{gue07a,
gue07b} and have found that the thinner the tachocline, the smaller
the intensity of the toroidal magnetic field at high latitudes (see
Fig. 7). A thin tachocline must be fully contained inside the
overshoot zone, in such a way that only part of the poloidal
magnetic field is able to reach it and then produce a small quantity
of toroidal field. The toroidal field generated there is not strong
enough to emerge. In Fig. 8, we compare the toroidal fields produced
for a thin tachocline with those produced with an intermediate one.
This result is also dependent of the magnetic diffusivity in the
convection zone, as it can be seen in Fig. 7, however, for a thin
tachocline the models always result weak toroidal fields above
$60^{\circ}$.

\begin{figure}[h!]
\begin{center}
\includegraphics[scale=0.7]{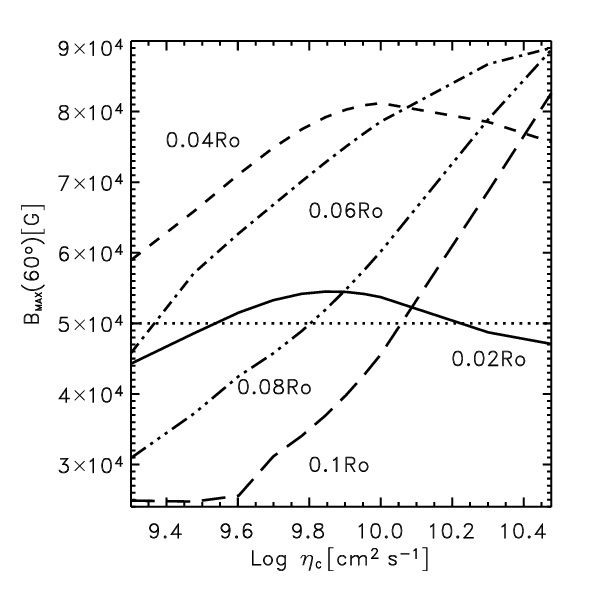}\\
\caption{Maximum of the toroidal magnetic field at the top of the
tacholine as a function of the diffusivity (in log-scale) at a
latitude of $60^{\circ}$. The different line styles represent
different widths of the tachocline $d_1$. The dotted line represents
the limit between buoyant and non-buoyant magnetic fields
$5\times10^4$ G. Only the values below this line will appear at the
desired latitudes {\it Extracted from
  \citet{gue07a}}.}
\label{fig7}
\end{center}
\end{figure}

\begin{figure*}[!t]
\includegraphics[scale=0.42]{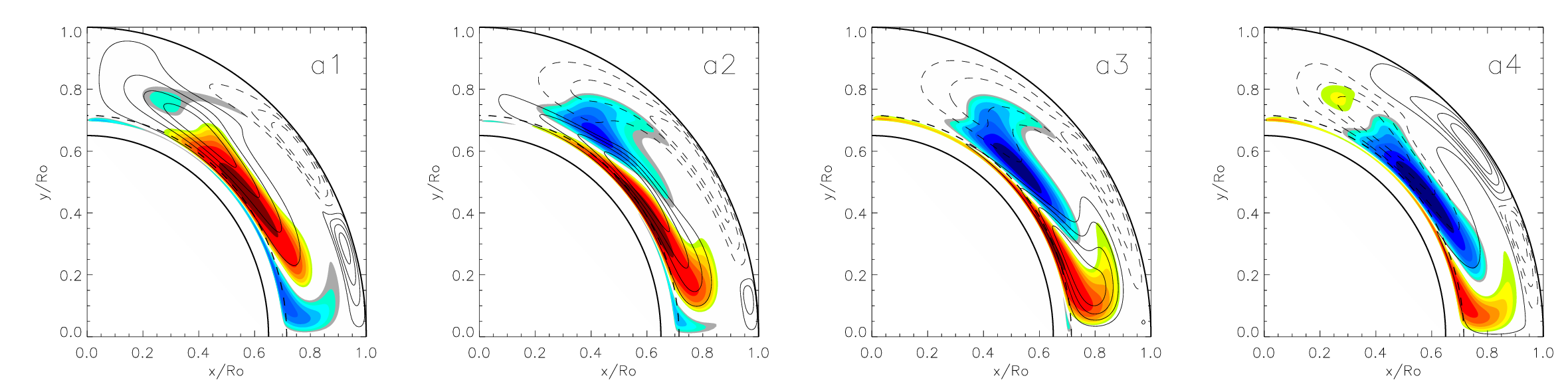}\\
\includegraphics[scale=0.42]{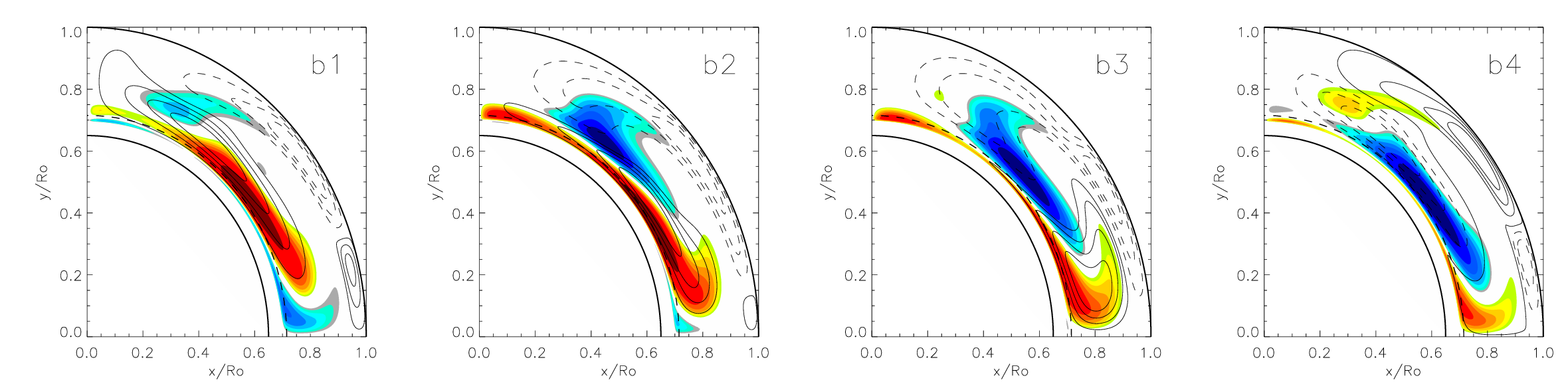}\\
\caption{Snapshots of the positive (negative) toroidal field contours
  are shown in blue (red) scale together with the field lines of the
  positive (negative) poloidal field in continuous (dashed) lines,
  for four different times ($T/8$, $T/4$, $3 T /8$ and $T/2$) along a
  half 11-year cycle. The upper (a) and middle
  (b) panels present the same snapshots for models with a thin ($0.02 \sr$)
  and an intermediate ($0.06 \sr$) tachocline {\it Extracted from
  \citet{gue07b}}.}
\label{fig8}
\end{figure*}

With regard to this question of the distribution of the toroidal
field, the role of the pumping is also important. While it pushes
poloidal fields inside the tachocline, it also pumps all the field
equatorward, in such a way that the toroidal fields formed inside
the convection zone due to latitudinal shear will go fast to the
stable region where they are stored and amplified before the
eruption \citep{gue08}.

In the case that we consider the sunspots as the product of toroidal
fields being formed at the near-surface layer, the latitude of
activity is easily explained by the pumping, as described in the
previous section  (see Fig. 6).

\section{The parity problem}

The anti-symmetric parity observed in the solar cycle is one of the
most challenging questions in the solar dynamo theory. The magnetic
parity in a model may depend on the location of the
$\alpha$ effect \citep{dikgil01,bon02}, or on the diffusive
coupling between the poloidal field in both hemispheres \citep{ccn04},
but it may also  be the result of the imprint of the quadrupolar form
of the meridional flow on the poloidal magnetic field, as argued by
\citet{char07}.  This could explain why models with the $\alpha$ effect
located at the upper layers (where the magnetic Reynols number and the
quadrupolar imprint of the meridional flow are larger) tend to
a quadrupolar parity faster than models with the $\alpha$ effect located
at the tachocline.

We have explored a little this problem in order to see whether the
turbulent pumping plays some role in it. We have found that models
without pumping result a quadrupolar solution. When beginning with a
dipolar initial condition, they spend  several thousand
 years
before switching to a quadrupolar solution (see Fig. 9a). This result
diverges from the one obtained by \citet{dikgil01} or \citet{ccn04} in
which the change begins only after around $500$ yr. This result
indicates the strong sensitivity of the parity to the initial
conditions,  in such a way that, for example, the present parity
observed in the sun could be temporary.

The models with full pumping (e.g., Fig. 5) conserve  the initial
parity either it is symmetric or anti-symmetric (see Fig
9b). This suggests that the strong quadrupolar imprint due to
meridional  circulation can be washed out when the full turbulent
pumping is  switched on.

On the other hand, the full pumping in models that include
near-surface shear tend to the dipolar parity from the first years
of integration (see continuous line of Fig. 9c). We explain this
result as a product of a better coupling between the poloidal fields
in both hemispheres (this coupling is due to the local $\alpha$
effect considered in these cases), plus the action of the pumping
eliminating the effect of the quadrupolar component of the
meridional flow on the poloidal magnetic field.

\begin{figure}[htb]
\centering
  \includegraphics[width=\columnwidth]{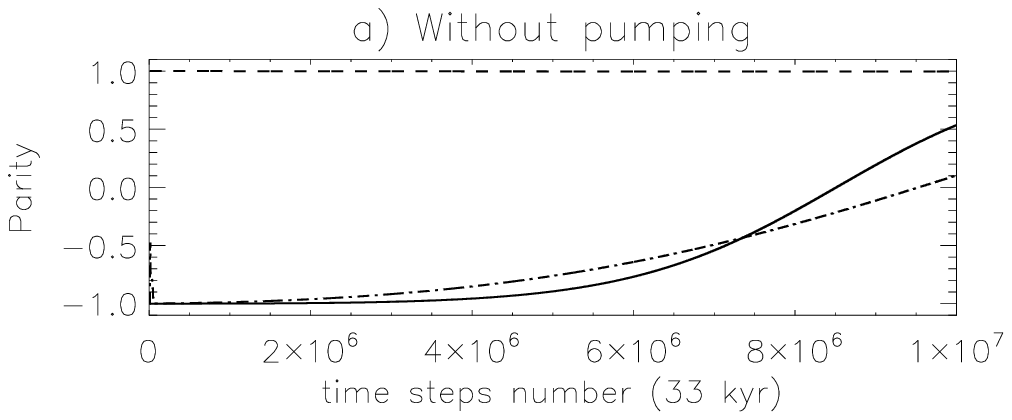}\\
  \includegraphics[width=\columnwidth]{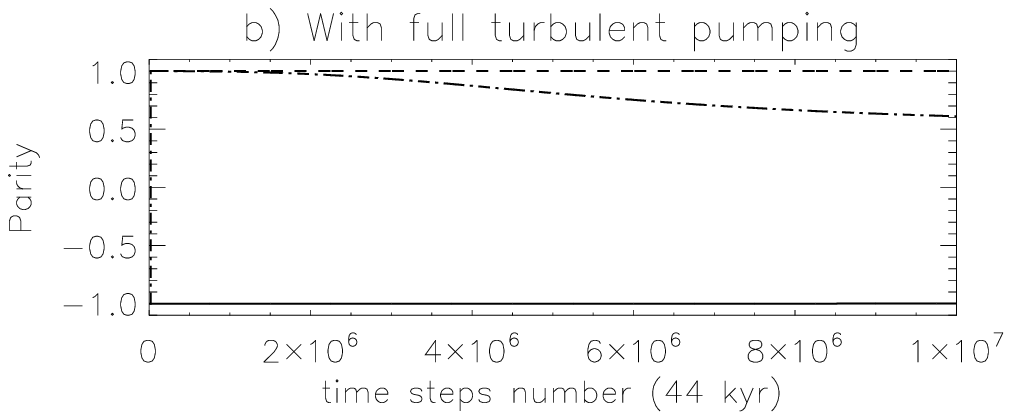}\\
  \includegraphics[width=\columnwidth]{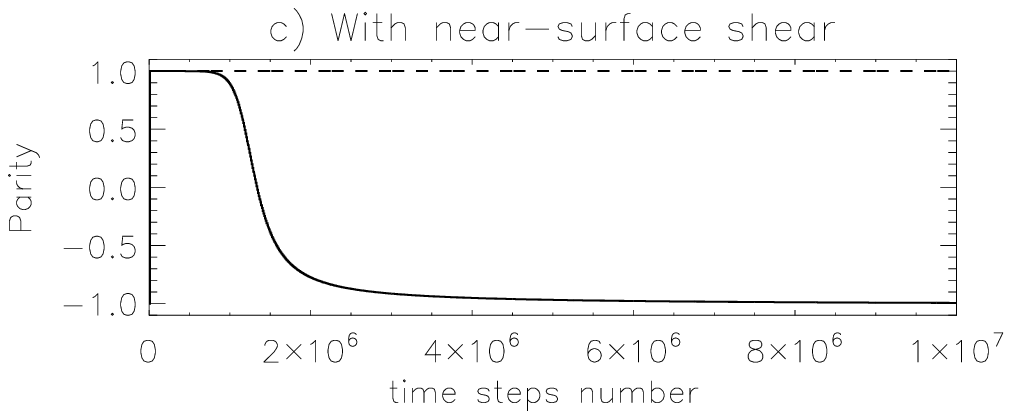}
\caption{Parity curves for the three classes of models
  considered, i.e., (a) for models without pumping (as, e.g., in
  Fig. 3); (b) for models with full pumping (as, e.g., in Fig. 5); and
  (c) for models with near-surface shear (as, e.g., in Figs. 4 and 6).
  In the panels (a) and (b), the continuous, dashed and dot-dashed
  lines correspond to symmetric, anti-symmetric and random initial
  conditions, respectively. In panel (c), the continuous
  line is used for the model with turbulent pumping while the
  dashed line is for the model without pumping {\it Extracted from
  \citet{gue08}}.}\label{fig8}
\end{figure}

\section{Summary and discussion}

We have used a mean field dynamo model in the BL approach in order
to look for answers to four current problems widely reported in the
literature of solar dynamo modeling.

Our results confirm the idea that there should be a magnetic layer
below the convection zone where magnetic fields which are mainly
produced in the convection zone are stored. We have found that it is
possible to have a flux-transport dynamo without a well defined
meridional flow pattern,  dominated by the pumping advection.
However, efforts in order to obtain a more realistic profile for the
meridional velocity profile are still required, as well as to obtain
a better comprehension of the real contribution of the pumping
velocities.

Nevertheless, numerical simulations including near-surface shear as
observed, provide also support for a near-surface magnetic layer, since
toroidal fields with intensities between $10^3$ - $10^4$ G are formed
there and since the radial differential rotation is negative at
lower latitudes, the direction of migration of the butterfly wings
constructed with these fields reproduces the one observed (e.g.,
Fig. 6).

The latitude of activity established by the observations (between
$\pm 30^{\circ}$) can be explained by both scenarios above for the
magnetic layer. For magnetic fields being stored in the overshoot
region, a thin ($\lesssim 0.2 \sr$) tachocline could be the solution
to avoid strong toroidal fields at the polar regions. On the other
hand, if the near-surface shear is considered, then the magnetic
pumping provides the required downwards flux of the weak polar
fields, letting only the toroidal fields to survive at the active
latitudes.

The parity in a dynamo solution is a problem that requires especial
attention. We have investigated this problem looking for the role
that the pumping could play in the solutions. Our simulations
support the idea that the quadrupolar solution in most of the models
is due to the strong quadrupolar imprint due to the one-cell
meridional flow pattern. This imprint is larger at the surface than
at the bottom of the convection zone, therefore the results of
\citet{dikgil01,bon02} which suggest that an $\alpha$ effect
operating at the tachocline results an anti-symmetric solution,
could be explained by this fact. We suggest that models with an
$\alpha$ effect operating close to the surface are also able to
generate anti-symmetric solutions, as observed, if a mechanism such
as pumping cleans the quadrupolar imprint.

More observational and theoretical efforts are necessary in order to
determine where the feet of the magnetic flux tubes responsible for
the sunspots are located. This is an issue that needs to be solved
before a more realistic coherent dynamo model can be constructed.

\bibliography{guerrero}

\end{document}